\documentclass[12pt]{article}

\usepackage{graphicx,amsmath,bm}

\begin{document}

\title{Low-lying $s=+1$ Pentaquark states in the Inherent Nodal Structure Analysis}

\author{{Yu-xin Liu$^{1,2,4,5}$, Jing-sheng Li$^{1}$, and Cheng-guang Bao$^{3,5}$}\\
\noalign{\vskip 5mm}
 \normalsize{$^{1}$ Department of Physics, Peking University,
Beijing 100871, China} \\
\normalsize {$^{2}$ Key Laboratory of Heavy Ion Physics,
Ministry of Education, Beijing 100871, China } \\
\normalsize{$^{3}$ Department of Physics, Zhongshan University,
Guangzhou 510275, China} \\
\normalsize{$^{4}$ Institute of Theoretical Physics, Chinese
Academy of Sciences, Beijing 100080, China} \\
\normalsize{$^{5}$ Center of Theoretical Nuclear Physics, National
Laboratory
of Heavy Ion Accelerator,}\\
\normalsize{Lanzhou 730000, China} }

%\date{\today}

\maketitle

\begin{abstract}
The strangeness $s=+1$ pentaquark states as $qqqq\bar{q}$ clusters
are investigated in this letter. Starting from the inherent
geometric symmetry, we analyzed the inherent nodal structure of
the system. As the nodeless states, the low-lying states are
picked out. Then the S-wave state $(J^P, T)= ({\frac{1}{2}}^{-},
0)$ and P-wave state $(J^P, T)= ({\frac{1}{2}}^{+}, 0)$ may be the
candidates of low-lying pentaquark states. By comparing the
accessibility of the two states and referring the presently
obtained K-N interaction potential, we propose that the quantum
numbers of the observed pentaquark state $\Theta^{+}$ may be
$(J^P, T)=({\frac{1}{2}}^{+}, 0)$ and $L=1$.
\end{abstract}



%\maketitle

\newpage

It has been proved that the quantum chromodynamics(QCD) is the
underlying theory for strong interaction, and the quark model is
remarkably successful in classifying the hadrons as composite
systems of quark and antiquark. Meanwhile the existence of
color-singlet multi-quark systems ($q^m\bar{q}^n$, $m+n>3$) has
also been predicted in QCD. This subject is very important, since
it is an appropriate place to investigate the quark behavior in
the short distance, which may shed a light on new physics.

Recently, LEPS\cite{LEPS}, DIANA\cite{DIANA}, CLAS\cite{CLAS},
SAPHIR\cite{SAPHIR} and many
others\cite{Neutrio03,CLAS032,HERMES03,SVD04,COSY04} reported that
they all observed a new resonance $\Theta$, with strangeness
$s=+1$. This is probably a 4-quark and 1-antiquark system. If it
is really a pentaquark state, it will be the first multi-quark
state observed in experiment. There have been then many
theoretical works to explain the properties of the $\Theta ^{+}$.
For example, the mass of $\Theta^{+}$ has been calculated in the
chiral soliton model\cite{DPP97,WM03,EKP04}, Skyrme
model\cite{Pras03}, diquark-triquark cluster
model\cite{KL03,KLV04,MSS04}, constituent quark
model\cite{Gloz03,SR03,Stancu04}, chiral SU(3) quark
model\cite{Zhang03}, QCD sum rules\cite{Zhu03,Math03,SDO03},
Lattice QCD\cite{Fodor03,Sasaki03,CH04}, large $N_c$
QCD\cite{CL03}, chiral perturbation approach{\cite{Ko03}, and so
on. Meanwhile, the general framework of
QCD\cite{JW031,Cap03,Carl03,JM03,SZ03} and the group theoretical
classification\cite{Wyb03,BGS03} have also been implemented to
explore the quantum numbers of the pentaquark state $\Theta^{+}$.
Nevertheless, the quantum numbers of the $\Theta ^{+}$ have not
yet been determined uniquely due to the dynamical model
dependence. Especially, the parity of the lowest-lying pentaquark
state is more controversial.  On the other hand, just based on the
symmetry of the intrinsic color-flavor space, many configurations
are possible for the pentaquark state (see, for example,
Ref.\cite{BGS03}). If one can fix which configuration is the
practical building block of the state, it will be much helpful to
assign the quantum numbers and to perform the numerical
calculation. If one can determine the quantum numbers (including
the parity) model independently, the result may be more reliable.
Fortunately, it has been known that the features of quantum states
depend on the distribution of wave-functions in the coordinate
space, more specifically, on the inherent nodal structure of the
wave-functions. The applications of the inherent nodal structure
analysis approach to six-nucleon and six-quark systems\cite{BL99,
LLB03} show that analyzing the inherent nodal structure can
uncover why a state with a specific set of quantum numbers is
lower or higher, and how the wave-function of a specific state is
distributed in the specific way in coordinate space. Then one can
pick out the real accessible configuration from the quite large
configuration space and fix the quantum numbers. We will thus take
the inherent nodal structure analysis approach to fix the quantum
numbers of the low-lying pentaquark states model independently in
this letter.

It has been well known that, in quantum physics, the point for the
wave-function to be zero is a node. The set of the variables for
the node to appear spans a nodal surface. Meanwhile, for usual
quantum system, the less nodes the configuration contain, the
lower energy the state has. For example, for the particle in an
infinite well with width $a$, the relation between the number $n$
of the nodes of the wave-function and the energy of the state is
$E_{n} = \frac{(n+1)^2 \pi ^2 \hbar ^2}{2 m a^2} $. For the
particle in a harmonic oscillating potential, the relation is
$E_{n} = (n + \frac{1}{2} ) \hbar \omega$. The nodal surface is
usually classified into the dynamical one and the inherent one.
The dynamical nodal surface depends on the dynamics of the system.
The inherent nodal surface (INS) relies on the inherent geometric
configuration of the system.

Let ${\cal{A}}$ be a geometric configuration in the
multi-dimensional coordinate space of a quantum system,
$\hat{O}_i$ be an element of the operation on wave function
$\Psi({\cal{A}})$ of the system, we have
$$\displaylines{ \hspace*{2cm}
\hat{O}_{i} \Psi({\cal{A}})= \Psi(\hat{O}_{i}{\cal{A}}) \, .
\hfill{(1)} \cr }$$
 If this set operation $\{\hat{O}_i \}$ (with $m$ elements)
leaves the configuration ${\cal{A}}$ invariant, i.e.,
$\displaystyle{\hat{O}_{i} {\cal{A}} = {\cal{A}}  \,}$,
 we have
$$\displaylines{\hspace*{2cm}
\hat{O}_{i} \Psi({\cal{A}})= \Psi({\cal{A}}) \, . \hfill{(2)} \cr
}$$
 Since the $\Psi({\cal{A}})$ spans a representation of
$\hat{O}_i$, Eq.~(2) can be written in a matrix form, which is in
fact a set of homogeneous linear algebraic equations. Therefore,
associated with the $m$ operators, there are $m$ sets of
homogeneous equations, that the $\Psi({\cal{A}})$ must obey.
However, a set of homogeneous equations does not always have
non-zero solutions. In the case that common non-zero solutions
fulfilling all the $m$ sets of equations do not exist, all
$\Psi({\cal{A}})$ must be zero at ${\cal{A}}$. Then a nodal
surface appears. Since such a nodal surface is determined by the
inherent geometric and intrinsic configuration but not by the
dynamics at all, it is referred to inherent nodal surface (INS).
It indicates that, when the particles form a shape with a specific
geometric symmetry, specific constraints are imposed on the
wave-function. Only the inherent nodeless components are the
accessible ones to the state. Then, with the inherent nodal
structure analysis we can pick out the main components from the
whole configuration space in a way independent of dynamics.

The wave function of the five-particle systems can usually be
written as a coupling of the orbital part and the internal part.
Since quark and antiquark are not identical to each other, the
wave-function is not antisymmetric via the interchange between
them. However, if we consider only the permutations among the 4
quarks, it should be antisymmetric, i.e. it has the symmetry
[1$^4$]. It has been well known that the $s=+1$ system with light
quarks possesses the internal symmetry
SU$_{CTS}$(12)$\supset$SU$_C$(3)$\otimes$SU$_T$(2)$\otimes$SU$_S$(2).
Let [f]$_O$, [f]$_C$ and [f]$_{TS}$ be the irreducible
representation(irrep) of the group associated with the orbital,
color and isospin-spin space, respectively, we shall have
$$ [1^4]\in[f]_O\otimes[f]_C\otimes[f]_{TS} \, . $$

The lack of direct experimental observation of free ``color
charge'' suggests that all the observable states should be SU(3)
color-singlets. We can thus restrict our study to the systems
whose 4 quarks have a color symmetry $[f]_{C} = [2 1 1]$. Then the
configuration of the $[f]_{O}$ and $[f]_{TS}$ can be fixed with
the group theoretical method. The obtained possible [f]$_{TS}$ and
[f]$_O$ are listed in Table I. It is obvious that such a orbital
and isospin-spin configuration space is very large. We should pick
out the important ones for the low-lying pentaquark state. As
mentioned above, we can do so by taking the inherent nodal
structure analysis approach.

\begin{table}[htbp]\begin{center}
\caption{\small{The irreducible representations of the
isospin-spin symmetry corresponding to each possible orbital
symmetry with color singlet restriction.}} \vspace*{5mm}
\begin{tabular}{l c l}
\hline \ $[f]_O$ &  & SU$_{TS}$(4) \\
\hline\ [4] &  & [3 1] \\
\ [3 1] &  & [4], [3 1], [2 2], [2 1 1] \ \\
\ [2 2] &  & [3 1], [2 1 1] \\
\ [2 1 1] &  & [3 1], [2 2], [2 1 1], [1$^4$] \ \\
\ [1$^4$] &  & [2 1 1] \\
\hline
\end{tabular}
\end{center}
\bigskip
\end{table}

It has been known that the wave-function of the state with total
angular momentum $J$, orbital parity $\pi$ and total isospin $T$
can be written as
 $$\displaylines{\hspace*{2cm}
\label{f1}\Psi=\sum_{L,S,\lambda} \Psi_{L\pi\lambda}^{S} \, ,
\hfill{(3)} \cr }$$
 with
$$\displaylines{\hspace*{2cm}
\label{f2}\Psi_{L\pi\lambda}^{S}=\sum_{i,M,M_{S}}C_{LM,SM_{S}}^{J',M_{J'}}
F_{LM}^{\pi\lambda_i}\chi_{TSM_S}^{\widetilde{\lambda}_i}  \, ,
\hfill{(4)} \cr } $$
 where $F_{LM}^{\pi\lambda_i}$ is a function of the spatial
coordinates, $\lambda$ denotes a representation of the $S_4$ group
(permutations among 4 quarks), and $i$ specifies a basis state of
this representation. The $L$, $S$ is the total orbital angular
momentum, the total spin of the 4 quarks, respectively. They are
coupled to $J'$ via the Clebsch-Gordan coefficients, and total
spin $J$ is formed by coupling the $J'$ and the antiquark's spin.
The $M$, $M_S$ and $M_{J'}$ are the Z-components of $L$, $S$ and
$J'$, respectively. The orbital parity of pentaquark is given by
$\pi=(-)^{L}$, and the total parity $P=-\pi=(-)^{L+1}$, since the
antiquark holds an instinct negative parity.
$\chi_{TSM_S}^{\widetilde{\lambda}_i}$ is a state in the
isospin-spin space with good quantum numbers $T$ and $S$, and
belonging to the $\widetilde{\lambda}$-representation, the
conjugate of $\lambda$. The $\lambda$ contained in $\Psi$ is
determined by $S$ and $T$. The result is listed in Table 2. Such a
$\Psi_{L\pi\lambda}^S$ is usually denoted as a $\lambda$-component
of $\Psi$.

\begin{table}[ht]\begin{center}
\caption{\small{Some of the allowed $\lambda$ and $[f]_{TS}$ for
isospin $T=0,1,2$ states of the four $u$- and $d$-quark system}}
\vspace*{5mm}
\begin{tabular} {|c|c|c|} \hline
$(S, T)$ & $[f]_{TS}$ &  $\lambda$ \\
\hline \raisebox{-2.0ex}[0cm][0cm] {(0, 0)} & [4] & [3 1]                              \\
\cline{2-3}   {}   & $[2 \; 2]$ & [3 1], [2 1 1]              \\
\hline \raisebox{-2.0ex}[0cm][0cm] {(1, 0)} & [3 1]   & [4], [3 1], [2 2], [2 1 1]      \\
\cline{2-3}   {}  & [2 1 1] & [3 1], [2 2], [2 1 1], $[1^4]$           \\
\hline \raisebox{-2.0ex}[0cm][0cm] {(2, 0)} & [3 1] & [4], [3 1], [2 2], [2 1 1]        \\
\cline{2-3}   {}  & [2 2] & [3 1], [2 1 ]                     \\
\hline \raisebox{-2.0ex}[0cm][0cm] {(0, 1)} & [3 1] & [4], [3 1], [2 2], [2 1 1]       \\
\cline{2-3}   {}   & [2 1 1] & [3 1], [2 2], [2 1 1], $[1^4]$          \\
\hline \raisebox{-2.0ex}[0cm][0cm] {(1, 1)} & [4], [3 1] & [4], [3 1], [2 2], [2 1 1]  \\
\cline{2-3}   {}   & [2 2]      & [3 1], [2 1 1]              \\
\hline (2, 1) & [3 1] & [4], [3 1], [2 2], [2 1 1]       \\
\hline \raisebox{-2.0ex}[0cm][0cm] {(0, 2)} & [3 1] & [4], [3 1], [2 2], [2 1 1]       \\
\cline{2-3}    {}  & [2 2] & [3 1], [2 1 1]                   \\
\hline (1, 2)& [3 1] & [4], [3 1], [2 2], [2 1 1]        \\
\hline (2, 2)& [4] & [3 1]                               \\
\hline
\end{tabular}
\end{center}
\bigskip
\end{table}

Let $i'$-$j'$-$k'$ be a body frame, the spatial wave-functions can
be expanded as
$$\displaylines{\hspace*{2cm}
F_{LM}^{\pi\lambda_i}(1234)=\sum_{Q}D_{QM}^{L}(-\gamma,-\beta,-\alpha)
F_{LQ}^{\pi\lambda_i}(1'2'3'4') \, , \hfill{(5)} \cr }$$
 where $D_{QM}^L$ is the Wigner function, $\alpha$, $\beta$ and
$\gamma$ are the Euler angles to specify the collective rotation,
$Q$ is the component of $L$ along $k'$, $(1234)$ and $(1'2'3'4')$
denote that the coordinates in $F_{LM}^{\pi\lambda_i}$ and
$F_{LQ}^{\pi\lambda_i}$ are related to the laboratory frame or to
the body frame, respectively. It turns out that the
$\{F_{LQ}^{\pi\lambda_i}\}$ span a representation of the rotation
group, space inversion group, and permutation group ($S_4$). Thus
the transformation property of the $F_{LQ}^{\pi\lambda_i}$ with
respect to the operations of the above groups is prescribed. This
fact will impose a very strong constraint on the
$F_{LQ}^{\pi\lambda_i}$, from which we can fix the practically
accessible configuration in orbital space.

Since the quarks are not identical to the antiquark, we can
consider only a special kind of configurations, in which the four
quarks form a geometric shape, and the antiquark locates at its
center due to the mechanical balance (analogous to that in the
diquark-antiquark model\cite{MSS04}). Considering the geometric
configuration of the four quarks, one can imagine that the
linkages among the quarks may form a tetrahedron, a tetragon, or
others. Recalling the lattice QCD result of the color flux-tube
structure of three-quark system\cite{IBSS03}, we know that there
exists genuine three-body interaction among the three quarks, and
the linkages between every two quarks may form a equilateral
triangle (the interaction is in Y shape). Extending such a
geometric feature to the four-quark system, we take the
equilateral tetrahedron (ETH) and the square into account in this
paper. In fact, as we shall discuss below, if the geometric
configuration is not so regular, less constraint is imposed to the
system.

For the geometric configuration in equilateral tetrahedron (ETH,
denoted also as ${\cal{A}}$ in the following), which is
illustrated in figure 1, we denote $O$ as the center of mass of
the four quarks (where the antiquark is located at), $O'$ as the
center between particles 1 and 2, $O''$ as the center between
particles 3 and 4, $r_{12}\bot k'$ and $r_{34}\bot k'$. Referring
to $R_{\delta}^{\vec{v}}$ as a rotation about the axis along the
vector $\vec{v}$ by an angle $\delta$, $p_{ij}$ as an interchange
of the particles i and j, $p_{ijk}$ as a permutation among the
particles i, j and k, $p_{ijkl}$ as a permutation among the
particles i, j, k and l, and $\hat{P}$ as a space inversion, we
know that the ETH is invariant to the operations
$$\displaylines{\hspace*{2cm}
\hat{O}_1 = p_{12}p_{34}R^{k'}_{\pi} \, , \hfill{(6)} \cr
\hspace*{2cm} \hat{O}_2 = p_{12}R^{i}_{\pi} \hat{P} \, ,
\hfill{(7)} \cr \hspace*{2cm} \hat{O}_3 =
p(1423)R^{k'}_{\frac{\pi}{2}} \hat{P} \, , \hfill{(8)} \cr
\hspace*{2cm} \hat{O}_4 = p(243) R^{n'}_{\frac{2\pi}{3}} \, .
\hfill{(9)} \cr }$$
 Inserting these to Eq.~(2) respectively, we have
$$\displaylines{\hspace*{1cm}
 \sum_{i'}\{g^{\lambda}_{ii'} [p_{12}p_{34}(-1)^Q - \delta_{ii'} ] \}
 F_{LQ}^{\pi\lambda_{i'}}({\cal{A}}) = 0 \, , \hfill{(10)} \cr \hspace*{1cm}
\sum_{i'}\{g^{\lambda}_{ii'} [p_{12} - \delta_{ii'} ] \}
F_{LQ}^{\pi\lambda_{i'}}({\cal{A}}) = 0 \, , \hfill{(11)} \cr
\hspace*{1cm} \sum_{i'}\{g^{\lambda}_{ii'} [p_{1423} (-i)^Q -
\delta_{ii'}  ] \} F_{LQ}^{\pi\lambda_{i'}}({\cal{A}}) = 0 \, ,
\hfill{(12)} \cr \hspace*{5mm} \sum_{i'Q'}\Big\{g^{\lambda}_{ii'}
\Big[ p_{243}\sum_{Q''}D_{Q''Q}^L(0,\theta,0)
e^{i\frac{2\pi}{3}Q''}D_{Q''Q'}^L(0,\theta,0) - \delta_{ii'}
\delta_{QQ'} \Big] \Big\} F_{LQ'}^{\pi\lambda_{i'}}({\cal{A}}) = 0
\, , \hfill{(13)} \cr }$$
 where  $\{ g^{\lambda}_{ii'} \}$ are the matrix elements of the
representation $\lambda$ and the $\theta$ with restriction
$\cos\theta=\sqrt{1/3}$.

Eqs.~(10)-(13) are the equations that the
$F^{\pi\lambda_i}_{LQ}({\cal{A}})$ have to fulfill. They are
homogeneous linear algebraic equations depending on $L$, $\pi$ and
$\lambda$. Because the rotational energy of the state with angular
momentum $L$ is $E_{r} \propto \frac{L(L+1)}{r^2}$, and the size
of quark system is very small, we take only the cases with $L < 2$
into account. Since the search for the non-zero solutions of the
homogeneous equations is trivial, we neglect describing the
evaluating process but list in Table 3 directly whether non-zero
solutions $F^{\pi \lambda _i}_{LQ}$ satisfying the above
constraints exist (marked with a letter ``A") at the configuration
$\lambda$ or not (marked with a letter ``$-$"). The table shows
obviously that, for only a few cases, there is a set of non-zero
solutions $F^{\pi\lambda_i}_{LQ}$ satisfying all these equations.
It implies that the associated $\lambda$-component
$\Psi^{\lambda}$ is non-zero at the ETH configurations. We may
then say that this $\lambda$-component is ETH-accessible. In other
cases, there are no non-zero solutions, all the $F^{\pi
\lambda_i}_{LQ}$ must be zero at the ETH configuration regardless
of its size and orientation. In such cases, the
$\lambda$-component is ETH-inaccessible.

\begin{table}[ht]\begin{center}
\caption{\small{The accessibility of the ETH and the square
configurations to the ($L \pi \lambda$) wave-functions.}}
\vspace*{5mm}
\begin{tabular} {|c|c|c|c| c| c| c|}
\hline       & $L^{\pi}$& [4]& $[3\, 1]$ & $[2\, 2]$ & $[2\, 1\, 1]$ & $[1^4]$ \\
\hline ETH    & 0$^+$ &   A  &   $-$   &  $-$  &  $-$    &   $-$     \\
\hline square & 0$^+$ &   A  &   $-$   &   A   &  $-$    &   $-$     \\
\hline ETH    & 1$^-$ &  $-$ &    A    &  $-$  &  $-$    &   $-$     \\
\hline square & 1$^-$ &  $-$  &   A    &  $-$  &   A     &   $-$     \\
\hline
\end{tabular}
\end{center}
\bigskip
\end{table}

Although the wave-functions are strongly constrained at the ETH,
they are less constrained in the neighborhood of the ETH. For
example, when the shape in Fig. 1 is prolonged along $k'$, which
can be called a prolonged tetrahedron, it is invariant to
$\hat{O}_1$, $\hat{O}_2$ and $\hat{O}_3$, but not $\hat{O}_4$.
Hence, the $F^{\pi\lambda_i}_{LQ}$ should fulfill the Eqs.~(10) to
(12). Evidently, a common non-zero solution of Eqs.~(10) to (13)
is necessarily a common solution of Eqs.~(10) to (12). Thus, if a
$\Psi^{\lambda}$ is non-zero at an ETH, it remains non-zero in its
neighborhood. In other words, an ETH-accessible component is
inherently nodeless in the domain surrounding the ETH.

For the configuration that the linkages among the four quarks form
a square (the antiquark locates at its center), as shown in
Fig.~2. It is evident that the square is invariant to
$$\displaylines{\hspace*{2cm}
\hat{O}_1' = p_{12}p_{34} \hat{P} \, , \hfill{(14)} \cr
\hspace*{2cm} \hat{O}_2' = R^{k'}_{\pi} \hat{P} \, , \hfill{(15)}
\cr \hspace*{2cm} \hat{O}_3' = p(34)R^{i'}_{\pi} \, , \hfill{(16)}
\cr \hspace*{2cm} \hat{O}_4' = p(1324) R^{k'}_{\frac{\pi}{2}} \, .
\hfill{(17)} \cr} $$
 These invariants lead also to constraints embodied in four sets of
homogeneous equations, and therefore the accessibility of the
square can be identified as listed in Table 3. As discussed above,
a square-accessible component is inherently nodeless in the domain
surrounding the square.

\begin{table}[ht]\begin{center}
\caption{\small{The predicted quantum numbers of the inherent
nodeless (low-lying) pentaquark states (with $L<2$) }}
\vspace*{5mm}
\begin{tabular} {p{1cm}|p{1cm} p{1cm} p{1cm} p{1cm} p{2.5cm} p{2cm}}
\hline state & \ $J^{P}$  & $L^{\pi}$ & $T$ & $S$ & $\,\;\;\;[f]_{TS}$ & \ \ \ \  $\lambda$ \\
\hline
 \ \ A &\ ${\frac{1}{2}}^+$ & $1^{-}$ & 0 & 0 & \ [4], [2 2] & \ \ [3 1]    \\
 \ \ B &\ ${\frac{1}{2}}^-$ & $0^{+}$ & 0 & 1 & \ \ \ \ [3 1] & \ \ \ [4]    \\
 \ \ C &\ ${\frac{1}{2}}^+$ & $1^{-}$ & 0 & 1 & [3 1], [2 1 1] & \ \ [3 1]    \\
 \ \ D &\ ${\frac{1}{2}}^-$ & $0^{+}$ & 1 & 0 & \ \ \ \ [3 1] & \ \ \ [4]    \\
\hline
\end{tabular}
\end{center}
\bigskip
\end{table}

Referring to Table 3, we find that, when a wave function
$\psi_{L\pi\lambda}$ possesses quantum numbers $(L^{\pi},
\lambda)=(0^{+}, [4])$, $(1^{-}, [3 \; 1])$, it can access both
the ETH and the square configurations. These and only these
$\psi_{L\pi\lambda}$ are inherently nodeless components in the two
important configurations and should be the dominant components of
the low-lying pentaquark states. We have then deduced four
possible low-lying states without taking any dynamical model. The
results can be listed in Table 4. All the other pentaquark states
should be remarkably higher in energy, because either they are
dominated by $L\geq 2$ components, or they do not contain
inherent-nodeless $\lambda$-components. In the case of $L=0$, two
states $(J^P,\; T)=(\frac{1}{2}^{-}, 0)$ and ($\frac{1}{2}^-$, 1)
(since the antiquark $\bar{s}$ is located at the center of the ETH
or the square, its total spin is just its intrinsic spin $1/2$)
contain both ETH-accessible and square-accessible components. They
are denoted as B and D in Table 4. These states, dominated by
component with spatial symmetry $[4]$ and isospin-spin symmetry
$[3 \; 1]$, are the low-lying S-wave states, while other S-wave
states must have much higher energies. In the case of $L=1$, Two
states with the same quantum numbers $(J^P, T)=(\frac{1}{2}^+, 0)$
contain both ETH-accessible and square-accessible components. As
the ones denoted as A and C in Table 4, they are associated with
the same spatial configuration $[3 \; 1]$ but different
isospin-spin symmetry $[f]_{TS}$. According to our analysis, these
two states are the low-lying ones with positive parity.

It is evident that if a state does not contain a collective
excitation of rotation, i.e. the angular momentum $L$ is zero, it
would be usually lower in energy than the state with $L > 0$. This
is particularly true for the systems with a very small size since
$E_{r} \propto L(L+1)/r^2$. It is then reasonable to assume that
the low-lying S-wave state B has an energy much lower than all the
low-lying P-wave states. It is thus the lowest pentaquark state.
{\it However}, referring to Table 4, one can easily recognize
that, the accessibility in the isospin-spin space for the S-wave
state B, D is only 1. Meanwhile the accessibility for both the
P-wave state A and C is 2. Then if the coupling in the
isospin-spin space can not be neglected, it is possible for the
P-wave state to appear as the lowest-lying state (i.e., with a
energy lower that of the S-wave state).

On the other hand, in view of the nucleon-meson collision, P-wave
resonance may also be important. If the S-wave state $(J^P, T) =
(\frac{1}{2}^{-}, 0)$ has an energy above the threshold of a
possible decay channel, this state can not be stable, the
pentaquark state will decay into a meson and a nucleon via strong
interaction. Furthermore, according to Table 2, a ``physical''
state $(S, T)=(1, 0)$ has four components associated with orbital
symmetry $[4]$, $[3\, 1]$, $[2\, 2]$, and $[2\, 1\, 1]$,
respectively, i.e. low-lying state B has three partners with
higher energy. This will probably lead to form a wide resonance,
which contradicts the experimental results. In the theoretical
point of view, the S-wave K-N potential in fall-apart mode makes
it very difficult to have a narrow width\cite{Zhang03,JM03}. Then,
the low-lying P-wave states, which possess a centrifugal barrier
to confine the nucleon and kaon in a narrow resonant state, may
become the stable ones instead. Comparing the P-wave states A and
C listed in Table 4, since the state A holds $(S, T)=(0, 0)$ and
unique orbital component $[3 \, 1]$, the state C has $(S, T) =
(1,0)$ and four orbital components $[4$, $[3 \, 1]$, $[2 \, 2]$
and $[2\, 1 \, 1]$, we propose that the P-wave state A, whose
orbital symmetry is uniquely $[3\; 1]$,  is the lowest-lying
stable state because the experimentally observed width is very
narrow. It is evident that such a result is consistent with many
of the previous predictions, for instance, the most original
chiral soliton model prediction\cite{DPP97}, recent Lattice QCD
result\cite{CH04}, Karliner and Lipkin's result\cite{KL03}, Stancu
and Risks's result\cite{SR03,Stancu04}, Jaffe and Wilczek's
result\cite{JW031,MSS04,JW032}, production cross section analysis
result\cite{NHK04}, and so on. By the way, considering the
spin-orbital coupling, we propose that the $J^{P} =
\frac{3}{2}^{+}$ state may also be the low-lying state.

In summary, with the inherent nodal structure being analyzed for
the system including four light quarks and one antiquark, we
propose dynamical model independently that the quantum numbers of
the lowest-lying pentaquark state $\Theta^{+}$ may be
$J^{P}=\frac{1}{2}^{+}$, $L=1$ and $T=0$. Such a result is
consistent with many previous predictions obtained  in concrete
dynamical models. Combining our model independent analysis and the
previous dynamical calculations, we would prefer to conclude that
the parity of the pentaquark state $\Theta^{+}$ is positive. Of
course, such a result is only a qualitative result. However,
taking the presently assigned configuration into dynamical model
calculations can obviously help to release the load of numerical
calculation. By the way, it is worth to mention that, if there
exists attractive interaction in the nucleon-kaon S-wave channel,
the parity of the $\Theta^{+}$ may be negative.

\bigskip

This work is supported by the National Natural Science Foundation
of China under the contract No. 19875001, 10075002, and 10135030
and the Major State Basic Research Development Program under
contract No.G2000077400. One of the authors (YXL) thanks the
support by the Foundation for University Key Teacher by the
Ministry of Education, China, too.

\newpage

\begin{figure}
\begin{center}
\includegraphics[scale=0.50,angle=0]{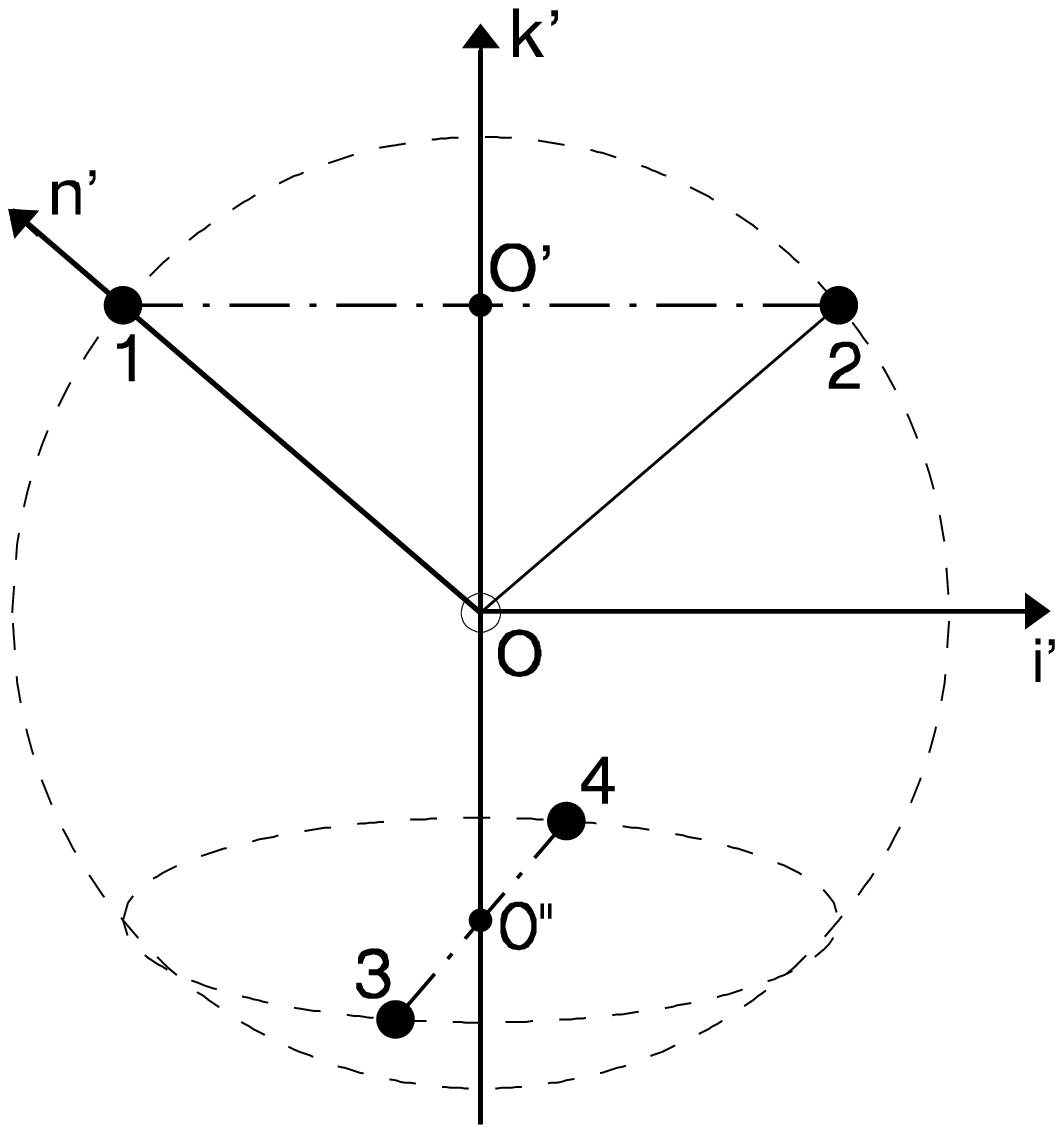}
\caption{A body frame for the ETH}
\end{center}
\end{figure}

\begin{figure}
\begin{center}
\includegraphics[scale=0.50,angle=0]{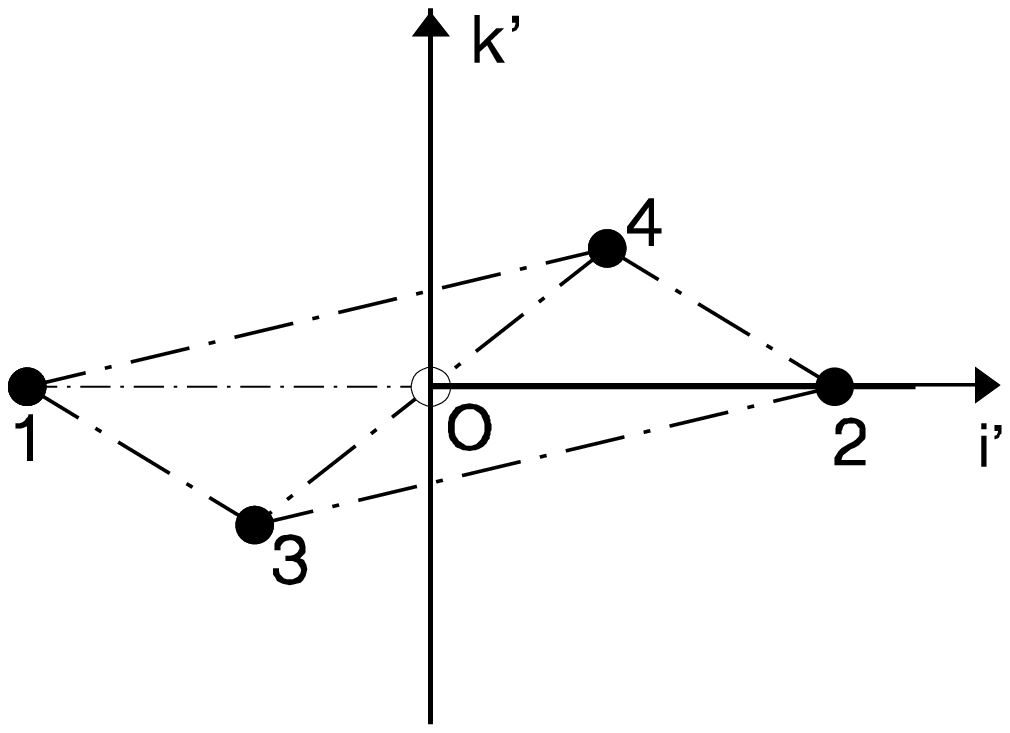}
\caption{A body frame for the square}
\end{center}
\end{figure}


\newpage


\begin{thebibliography}{50}

\bibitem{LEPS} T. Nakano \textit{et.al.}, Phys.Rev.Lett. {\bf 91} (2003)
012002.

\bibitem{DIANA} V.V. Barmin \textit{et.al.}, Phys.Atom.Nucl. {\bf 66} (2003)
1715.

\bibitem{CLAS} S. Stepanyan \textit{et.al.}, arXiv:
hep-ex/0307018.

\bibitem{SAPHIR} J. Barth \textit{et.al.}, arXiv: hep-ex/0307083.

\bibitem{Neutrio03} A. E. Aratayn, A. G. Dololenko, and M. A. Kubantsev, arXiv: hep-ex/0309042.

\bibitem{CLAS032} V. Kubarovsky, {\it et al.}, arXiv: hep-ex/0311046.
\bibitem{HERMES03} A. Airapetian, {\it et al.}, arXiv: hep-ex/0312044.
\bibitem{SVD04} A. Aleev, {\it et al.}, arXiv: hep-ex/0401024.
\bibitem{COSY04} M. Abdel-Bary, {\it et al.}, arXiv: hep-ex/0403011.

\bibitem{DPP97} D. Diakonov, V. Petrov, M. Polyakov, Z.Phys. {\bf A
359} (1997) 305.

\bibitem{WM03} B. Wu, and B. Q. Ma, arXiv: hep-ph/0311331.

\bibitem{EKP04} J. Ellis, M. Karliner, and M. Prasza\'{l}owicz, arXiv: hep-ph/0401127.

\bibitem{Pras03} M. Prasza\'{l}owicz, Phys. Lett. {\bf B 575} (2003) 234.

\bibitem{KL03} M. Karliner, and H. J. Lipkin, Phys. Lett. {\bf B 575} (2003) 249;
{\it ibid}, arXiv: hep-ph/0402260.

\bibitem{KLV04} N. I. Kochelev, H. J. Lee, and V. Vento, arXiv: hep-ph/0404065.

\bibitem{MSS04} D. Melikhov, S. Simula, and B. Stech, arXiv: hep-ph/0405037.

\bibitem{Gloz03} L. Ya Glozman, Phys. Lett. {\bf B 575} (2003) 18.

\bibitem{SR03} Fl. Stancu, and D.O.Riska, Phys. Lett. {\bf B 575} (2003) 242.

\bibitem{Stancu04} Fl. Stancu, arXiv: hep-ph/0402044.

\bibitem{Zhang03} F. Huang, Z. Y. Zhang, Y. W. Yu, and B. S. Zou, Phys. Lett. {\bf B 586} (2004) 69.

\bibitem{Zhu03} S. L. Zhu, Phys. Rev. Lett. {\bf 91} (2003) 232002.

\bibitem{Math03} R.D. Matheus, F.S. Navarra, M. Nielson, R. Rodrigues da Silva,
and S. H. Lee, Phys. Lett. {\bf B 578} (2004) 323..

\bibitem{SDO03} J. Sugioyama, T. Doi, and M. Oka, Phys. Lett. {\bf B 581} (2004) 167.

\bibitem{Fodor03} F. Csikor, Z. Fodor, S. D. Katz, and T. G. Kov\'{a}cs,
J. High Energy Phys. 0311 (2003) 070.

\bibitem{Sasaki03} S. Sasaki, arXiv: hep-lat/0310014.

\bibitem{CH04} T. W. Chiu, and T. H. Hsieh, arXiv: hep-ph/0403020.

\bibitem{CL03} T. D. Cohen, R. F. Lebed, Phys. Lett. {\bf B 578} (2004) 150.

\bibitem{Ko03} P. Ko, J. Lee, T. Lee, and J. Park, arXiv: hep-ph/0312147.

\bibitem{JW031} R. Jaffe, and F. Wilczek, Phys. Rev. Lett. {\bf 91} (2003) 232003.

\bibitem{Cap03} S.Capstick, P.R.Page, W.Roberts, Phys.Lett. {\bf B
570} (2003) 185; P. R. Page, arXiv: hep-ph/0310200.

\bibitem{Carl03} C. E. Carlson, C. D. Carone, H. J. Kwee, V. Nazaryan, Phys. Lett.
{\bf B 573} (2003) 101.

\bibitem{JM03} B. K. Jennings, and K. Maltman, arXiv: hep-ph/0308286.

\bibitem{SZ03} E. Shuryak, and I. Zahed, arXiv: hep-ph/0310270.

\bibitem{Wyb03} B. G. Wybourne, arXiv: hep-ph/0307170.

\bibitem{BGS03} R. Bijker, M.M. Giannini, E. Santopinto,
arXiv: hep-ph/0310281.

\bibitem{BL99} C. G. Bao, and Y. X. Liu, Phys. Rev. Lett. {\bf 82} (1999) 61.

\bibitem{LLB03} Y. X. Liu, J. S. Li, and C. G. Bao, Phys. Lett. {\bf B 544} (2002)
280; Phys. Rev. {\bf C 67} (2003) 055207; Mod. Phys. Lett. {\bf A
18} (2003) 414.

\bibitem{IBSS03} H. Ichie, V. Bornyakov, T. Streuer, and G.
Schierholz, Nucl. Phys. {\bf A 721} (2003) 899.

\bibitem{JW032} R. Jaffe, and F. Wilczek, arXiv: hep-ph/0312369.

\bibitem{NHK04} S. I. Nam, A. Hosaka, and H. C. Kim, arXiv:
hep-ph/0403009.

\end{thebibliography}
\end{document}